\RequirePackage{ifpdf}
\documentclass[12pt,letterpaper]{article}
\pdfoutput=1
\usepackage{jheppub}
\usepackage{epsfig}
\usepackage{enumitem}
\usepackage[latin1]{inputenc}
\usepackage{bbm,amsfonts}
\usepackage{rotating,graphicx}
\usepackage{amssymb,amsmath,amsfonts,mathtools}
\usepackage{fancybox,diagbox}
\usepackage{enumerate}
\usepackage{dsfont}
\usepackage{verbatim}
\usepackage{wrapfig}
\usepackage{slashed}
\usepackage{shuffle}
\usepackage{subcaption}
\usepackage{braket}
\usepackage{pdflscape}
\usepackage{accents}
\usepackage{afterpage}

\author[a]{Marco S. Bianchi}
 
\affiliation[a]{Niels Bohr Institute, University of Copenhagen, Blegdamsvej 17, 2100 Copenhagen $\emptyset$,\\
Denmark}

\emailAdd{marco.bianchi@nbi.ku.dk}  
  
\abstract{I consider three-point functions of one protected and two unprotected twist-two operators with spin in ${\cal N}=4$ SYM at weak coupling. At one loop I formulate an empiric conjecture for the dependence of the corresponding structure constants on the spins of the operators.
Using such an ansatz and some input from explicit perturbative results, I fix completely various infinite sets of one-loop structure constants of these three-point functions.
Finally, I determine the two-loop corrections to the structure constants for a few fixed values of the spins of the operators.}

\title{On structure constants with two spinning twist-two operators} 

\newcommand{\be}{\begin{equation}}
\newcommand{\ee}{\end{equation}}
\newcommand{\beq}{\begin{equation}}
\newcommand{\eeq}{\end{equation}}
\newcommand{\bea}{\begin{eqnarray}}
\newcommand{\eea}{\end{eqnarray}}
\newcommand{\ena}{\end{eqnarray}}

\def\Tr{\textrm{Tr}}

\numberwithin{equation}{section}

\def\clock{{\count0=\time
           \divide\count0 60
           \ifnum\count0<10 0\fi\the\count0
           \multiply\count0 -60 \advance\count0 \time
           :\ifnum\count0<10 0\fi \the\count0
         }}
\newcommand{\timestamp}{{\small\vbox{\hbox{\tt\jobname.tex}
\hbox{\the\day/\the\month/\the\year, \clock}}}}
\newlength{\dhatheight}
\newcommand{\doublehat}[1]{%
    \settoheight{\dhatheight}{\ensuremath{\hat{#1}}}%
    \addtolength{\dhatheight}{-0.35ex}%
    \hat{\vphantom{\rule{1pt}{\dhatheight}}%
    \smash{\hat{#1}}}}

\begin{document}

\maketitle
\allowdisplaybreaks

\section{Introduction}

The computation of three-point correlation functions of local operators in ${\cal N}=4$ SYM has been recently boosted by the hexagon approach \cite{Basso:2015zoa}, based on the integrability of the model \cite{Beisert:2010jr}. This proposal has stimulated various impressive perturbative tests at weak coupling at rather high loop order \cite{Eden:2015ija,Basso:2015eqa,Eden:2016aqo,Goncalves:2016vir,Basso:2017muf,Georgoudis:2017meq} and recipes for extending it to compute higher-point correlators \cite{Eden:2016xvg,Fleury:2016ykk,Fleury:2017eph,Basso:2017khq,Chicherin:2018avq,Coronado:2018ypq}. 
Besides, this has provided a framework where integrability is capable of determining color subleading effects \cite{Bargheer:2017nne,Eden:2017ozn,Bargheer:2018jvq}, which had long been elusive within integrability.

In light of such developments, in this paper I tackle the problem of computing perturbatively three-point functions of twist-two operators in ${\cal N}=4$ SYM. In particular, I analyze the case where two of such operators are unprotected and the third is BPS.
The main focus is on unprotected twist-two operators with spin, belonging to the $sl(2)$ sector of the theory.
As an aside, I also provide results for the somewhat simpler case where one of these operators is replaced by a scalar Konishi.
The target is the determination of the structure constants up to two-loop order.

In order to do so, I apply a method proposed in \cite{Plefka:2012rd} and that I recently developed further in \cite{Bianchi:2018zal}.
It basically consists in considering a soft limit for one of the operators, projecting the three-point function onto a two-point problem, which is much simpler to handle.
The structure constant is then retrieved by comparison with the expected structure of three-point functions, which is fixed in conformal field theories.
There are some subtleties connected to regularization, which could in principle undermine the success of such a procedure and in particular the comparison to extract the structure constant. I discuss them and provide a recipe for overcoming these technical difficulties in the problem under consideration.
This basically guarantees that the method can be consistently applied to compute the three-point functions I am after.

Still, the presence of two operators with spin complicates the calculation considerably. On the one hand, these three-point functions involve an increasing number of independent structure constants, corresponding to the various tensor structures that satisfy conformal Ward identities separately.
On the other hand, the relevant diagrams and Feynman integrals are rather challenging to evaluate in full generality and I could only solve them in a case-by-case analysis at fixed spin. Clearly, the computational effort required by it increases with the spins (and the perturbative order, obviously). 
Hence, this method might not be the most efficient for performing such a computation, but I wanted to probe how far I could push such a direct approach and to gather some sound perturbative data, before proceeding with more sophisticated techniques.

At one loop, a multitude of structure constants can be determined in a reasonable amount of time.
This has allowed me to formulate and test a conjecture for the dependence of the structure constants on the spins of the operators and supplement it with some explicit results.
Such an ansatz depends on a number of undetermined parameters, whose number grows linearly according to the tensor structure it corresponds to, in terms of a counting that I explain in the main body.
I fixed them in a few cases by direct perturbative computation, thereby providing some infinite sets of structure constants (conjecturally), that in particular determine completely the three-point functions in the case of one operator up to spin 14 and another with generic spin.
At two loops I just table the results for a few structure constants at fixed, and sufficiently low, values of the spins of the operators.

I conclude with some perspectives on future avenues of research. It would be interesting to compare these perturbative results with other approaches to three-point functions, in particular, the recent developments in the realm of integrability, that I mentioned at the beginning.
Also, three-point functions emerge in the OPE analysis of higher-point correlators and re-deriving and hopefully overperforming the results presented here with such a method is a stimulating idea.
In particular, this might conceivably shed more light or complete the conjecture at one loop, that I referred to above. Such an analysis should be doable, perhaps building on previous results on higher-point functions at one loop \cite{Drukker:2008pi} and the recent progress granted by integrability and hexagonalization \cite{Eden:2016xvg,Fleury:2016ykk}, but I have not attempted it yet.

The plan of the paper is as follows. In the next section I present the operators explicitly and recall the properties of their two- and three-point functions.
Then I outline the method that I used for the computation, which, in a nutshell, consists in forcing a vanishing momentum limit for one of the operators  \cite{Plefka:2012rd}.
Two further sections follow in which I spell out the results at one and two loops respectively.

\section{The operators and their two- and three-point functions}\label{sec:operators}

The main ingredients of this study are twist-two operators in ${\cal N}=4$ SYM. 
They are constructed out of the complex scalars of the model $X_a$ ($a=1,2,3$ or $(X,Y,Z)$) and a symmetric and traceless application of covariant derivatives $D$ and have the schematic form
\begin{equation} 
O^j_{ab} \equiv \Tr (D^{k}X_a D^{j-k} X_b) + \dots
\end{equation}
I will focus mainly on the operators in the $sl(2)$ sector with all indices projected on the light-cone, via contraction with null vectors $z_i$. I will consider two distinct such operators and therefore allow for a pair of null vectors $z_1$ and $z_2$ and denote their scalar product as $z_{12}\equiv z_1 \cdot z_2$. In two- and three-point functions I shall clearly consider the complex conjugate operators as well, which I shall denote with bars.
The explicit expression of the operators at a given spin $j$ is provided in terms of the coefficients of Gegenbauer polynomials $C_j^\nu$
\begin{equation}\label{eq:twist2}
\hat O_j = \sum_{k=0}^j\, a_{jk}^{\frac{d-3}{2}}\, \Tr\left( \hat D^k X \hat D^{j-k} X \right) \qquad,\qquad \hat D = D_{\mu} z_1^{\mu}
\end{equation}
where
\begin{equation}
\sum_{k=0}^j\, a_{jk}^{\nu}\, x^k y^{j-k} = (x+y)^j\, C_j^{\frac{d-3}{2}}\left( \frac{x-y}{x+y} \right)
\end{equation}
and $d$ is the space-time dimension.
In the notation above I have indicated a contraction with the $z_1$ vector by a hat and I shall adopt a double hat for projection on $z_2$, likewise. At tree level, the covariant derivatives are effectively simple derivatives, but at higher orders in perturbation theory they include diagrammatic corrections involving the gauge field of the theory.

At spin 0 the operators collapse onto chiral primary BPS objects, whose dimension does not receive quantum mechanical corrections. On the contrary, the operators with non-vanishing spin are unprotected and possess anomalous dimensions, which were determined perturbatively at high orders \cite{Kotikov:2003fb,Kotikov:2004er} and which have played a pivotal role in the development of AdS/CFT integrability \cite{Staudacher:2004tk,Eden:2006rx,Belitsky:2006av,Beisert:2006ez}.

I shall also consider a twist-two scalar unprotected operator with a famous anomalous dimension computed to high orders \cite{Velizhanin:2008rw}, which is the Konishi R-symmetry singlet
\begin{equation}\label{eq:konishi}
K \equiv \Tr (X_a \bar X^a) 
\end{equation}
with summation over flavor indices understood.

\paragraph{Two-point functions}

The two-point functions of operators with spin in a conformal theory, such as ${\cal N}=4$ SYM, are constrained to be non-vanishing only for operators of the same spin and dimension.
In general, these two-point functions suffer from ultraviolet divergences and the corresponding operators have to be re-normalized.
In particular, the twist-two operators which I consider here only mix with descendants of lower spin operators with the same total spin $j$, such as $\partial^k O_{j-k}$. The re-normalized field is defined as
\begin{equation}\label{eq:renormalization}
\hat {\cal O}_j = \sum_{k=0}^{j} Z_{j,k}\, \hat \partial^k \hat O_{j-k} = Z_j\, \hat O_j + \sum_{k=1}^{j} B_{j,k}\, \hat \partial^k \hat O_{j-k}
\end{equation}
where in the last equality I have separated the diagonal part from the off-diagonal mixing terms.
Incidentally, for odd spins the corresponding operators are descendants themselves. Hence, I will consider even spins $j_1$ and $j_2$ from now on, limiting the discussion to primaries.
The mixing pattern has been studied explicitly for instance in \cite{Belitsky:2007jp}, from which I use some explicit results in section \ref{sec:1loop}. Moreover, the spectral problem for such operators has been successfully addressed using the conjectured integrability of ${\cal N}=4$ SYM \cite{Beisert:2010jr}.

After re-normalizing the operators and finding the eigenstates of the dilatation operator, their two-point function is completely determined by conformal symmetry up to an overall normalization, which is just the normalization of the operators themselves
\begin{equation}\label{eq:2point}
\left\langle \hat {\cal O}^j(0) \doublehat{\bar{\mathcal{O}}} ^k(x) 
\right\rangle = C(g^2,N)\, \delta^{jk} \frac{\hat I^j}{|x|^{2\Delta}}
\end{equation}
where $\Delta$ is the full dimension of the operator, inclusive of quantum corrections, and $j$ is its spin. The tensor structure is encoded by the object
\begin{equation}\label{eq:Itensor}
\hat I \equiv I_{\mu\nu}\, z_1^{\mu}\, z_2^{\nu} \qquad, \qquad I_{\mu\nu} \equiv \eta_{\mu\nu} - 2\, \frac{x_{\mu}x_{\nu}}{x^2}
\end{equation}
contracted with two in principle distinct null vectors $z_1$ and $z_2$ for the two operators (in practical computations of two-point functions I use the same, since indeed the structure is completely fixed and I will only be after the normalization and its quantum corrections).
The normalization can then be fixed canonically in such a way to obtain an orthonormal set of operators, according to \eqref{eq:2point}.

\paragraph{Three-point functions}

The functional form of three-point functions is likewise fixed by conformal symmetry.
I will be mainly interested in the case of two primary operators with spin and a third scalar.
In such a situation a properly transforming tensor structure can be constructed out of combinations of the tensor $I$ \eqref{eq:Itensor} contracted with various pairs of indices. The size of such a combination increases with the spins and each independent structure can appear in the three-point function with an arbitrary coefficient. Therefore, several structure constants exist in general.
In particular, I will consider three-point functions of a twist-two operator ${\cal O}_{j_1}$ of spin $j_1$, another operator (or its conjugate, more precisely) with spin $j_2$ and a scalar BPS operator, with suitable flavor indices so as to allow for a tree-level contraction.
The corresponding three-point function possesses the form \cite{Sotkov:1976xe}
\begin{equation}\label{eq:3ptstructure}
\left\langle \hat {\cal O}_{j_1}(x_1)\, \, {\cal O}_{BPS}(x_2) \,\,\doublehat{{\cal O}}_{j_2}(x_3) \right\rangle = \frac{ \displaystyle
\sum_{l=0}^{\text{min}(j_1,j_2)}\, {\cal C}_l\, \frac{\hat Y_{32,1}^{j_1-l}\, \doublehat{ Y}_{12,3}^{j_2-l}}{\left(x_{13}^2\right)^l} \left( z_{12} - 2\, \frac{\hat x_{13}\, \doublehat{x}_{13}}{x_{13}^2} \right)^l}{|x_{12}|^{\Delta_{12,3}-j_1+j_2} |x_{23}|^{\Delta_{23,1}+j_1-j_2} |x_{13}|^{\Delta_{31,2}-j_1-j_2}}
\end{equation}
where I define
\begin{equation}
\hat Y_{ij,k} \equiv Y_{ij,k}^{\mu}\, z_{1\,\mu}\qquad,\qquad Y_{ij,k}^\mu \equiv \frac{x_{ik}^\mu}{x_{ik}^2} - \frac{x_{jk}^\mu}{x_{jk}^2}
\end{equation}
and I am using the shorthand notation
\begin{equation}
x_{ij} \equiv x_i - x_j \qquad,\qquad \Delta_{ij,k} \equiv \Delta_i+\Delta_j-\Delta_k
\end{equation}
In particular, assuming from now on that $j_1\geq j_2$, there are $j_2+1$ independent structure constants ${\cal C}_l$ to be computed.
Accordingly, if $j_2=0$ only one structure constant survives.
These are in general functions of the coupling $g^2$ (and the rank of the gauge group $N$) or the t' Hooft coupling $\lambda$ (in this note I will be using the 't Hooft coupling for compactness, as there are no color-subleading corrections up to the perturbative order I will be working at)
\begin{equation}
{\cal C}_l = {\cal C}_l^{(0)} + {\cal C}_l^{(1)}\, \lambda + {\cal C}_l^{(2)}\, \, \lambda^2 + \dots
\end{equation}
and the problem that I address in this paper is their computation at one- and two-loop order at weak coupling.

As a final remark, the structure constants depend of course on the normalizations of the operators, therefore I fix them to be orthonormal as explained above and I will also often consider the ratio of the quantum corrections of the structure constants by their tree-level counterpart.

\section{The perturbative computation}\label{sec:method}

I work in ${\cal N}=4$ SYM with $SU(N)$ gauge group. The results presented in this article, that is up to second order in perturbation theory, do not exhibit any non-planar correction. Therefore I will use ubiquitously the 't Hooft coupling $\lambda =\frac{g^2 N}{16 \pi ^2}$ as the perturbative expansion parameter ($g$ is the Yang-Mills coupling constant).

In order to fix the structure constants I apply the method of integrating over space-time one of the insertion points of the operators in the three-point function, thereby reducing it effectively to a two-point correlator problem.  The idea consists in performing a limit where the computation simplifies considerably, but which still ensures to recover the complete information on the structure constants.
This is possible since the overall structure of the three-point functions is fixed by conformal symmetry, as recalled above. From the comparison between the integrated form of the general structure of the three-point function \eqref{eq:3ptstructure} and the actual perturbative computation, one can extract the desired structure constants.
This method has been applied previously in \cite{Plefka:2012rd} and extended more recently in \cite{Bianchi:2018zal}, where more technical details can be found.

For the crucial step concerning the comparison between the integrated quantities, I stress that such a computation is plagued by divergences, coming both from the UV properties of the unprotected operators, which are eventually re-normalized away, and from IR singularities introduced by the additional integration. This, in fact, corresponds in momentum space to a soft limit for the given operator. These divergences require a regulator, which I choose to be dimensional regularization. In particular, I use the dimensional reduction scheme, so as to preserve supersymmetry and enforce a vanishing perturbative $\beta$-function.
This allows to carry out the relevant perturbative computation retaining a dependence on the dimensional regulator $\epsilon$.

A pivotal point for the success of the procedure, namely for being able to reconstruct the original structure constants by comparison, is in fact connected to regularization.
Indeed, while the perturbative computation is consistently performed in $d=4-2\epsilon$ dimensions, the general expression for three-point functions in conformal field theories \eqref{eq:3ptstructure} is derived using conformal symmetry in strictly integral dimension.
Such a detail could therefore introduce a fatal mismatch, potentially jeopardizing the whole derivation.
Still, as highlighted in \cite{Bianchi:2018zal}, in certain cases the integration procedure of the generic three-point functions turns out to be insensitive (up to the desired finite order in $\epsilon$ for the structure constants) to the $\epsilon$ fine-print.
Such a situation arises naturally when the integration is completely finite. This is for instance the case of the computation in \cite{Plefka:2012rd}, where three-point functions with only one operator with spin were considered and the integration point was chosen to be that corresponding to such an operator. In \cite{Bianchi:2018zal} I remarked that even if the integration introduces a divergence there are still special cases where a prediction can be extracted consistently.

This is important for three-point functions with two-operators with spin.
In this situation, integrating \eqref{eq:3ptstructure} over one of the insertion points would yield in general a polynomial (up to a common denominator) in the Lorentz invariant $z_{12}$, whose coefficients possess fixed powers of the distance between the unintegrated points, contracted with $z_1$ and $z_2$. Such coefficients also mix the various structure constants appearing in \eqref{eq:3ptstructure} and the particular combination depends on which point is integrated.

In particular, integrating over one of the operators with spin does not yield a finite result any longer, at a difference with respect to the case of three-point functions with only one spinning operator \cite{Plefka:2012rd}. On the contrary, it introduces divergences that spoil a direct comparison of the structure constants completely. More precisely, only one of the coefficients of the various powers of $z_{12}$ arising is finite and meaningful, but the others are not. In turn, this prevents from extracting information on the structure constants because it provides only one equation for three or more unknowns.

Choosing the integration point to be at the insertion of the operator without spin produces in general a divergent answer again. Still, the result allows for a direct comparison of the structure constants, provided the operator has vanishing anomalous dimension, i.e.~that it is protected. This occurs thanks to the fact that the integrated three-point function presents only simple poles in $\epsilon$ at any perturbative order (that arise from the tree-level space-time structure of the correlator), whose residue is proportional to a combination of structure constants. This differs from the case mentioned above, where the order of poles in $\epsilon$ increases with the perturbative order and the determination of structure constants at a given loop would require the knowledge of the precise dependence of the three-point functions on $\epsilon$ at loop level.
In particular, for the case at hand, integrating over the position of the BPS operator, which I chose to be at $x_2$, I find
\begin{align}\label{eq:integrated}
& \int d^{4-2\epsilon}x_2\,\, \left\langle \hat {\cal O}_{j_1}(x_1)\, \, {\cal O}_{BPS}(x_2) \,\,\doublehat{{\cal O}}_{j_2}(x_3) \right\rangle = 
\\&~~~~~
=-\frac{\pi^2\, \hat x_{13}^{\, j_1}\, \doublehat{x}_{13}^{\,j_2}}{\epsilon\, \left(x_{13}^2\right)^{1+j_1+j_2}}\, \sum_{l=0}^{\text{min}(j_1,j_2)}\, \sum_{k=l}^{\text{min}(j_1,j_2)}\, (-2)^{k-l} \left(\begin{array}{c}k\\ l\end{array}  \right)\, {\cal C}_k\, \left(\frac{x_{13}^2\, z_{12}}{\hat x_{13}\, \doublehat{x}_{13}}\right)^l + O(\epsilon^0)\nonumber
\end{align}
This is the key technical point which allows for the computation of the structure constants in the case of two operators with spin. Remarkably, the aforementioned equation arising when integrating on the insertion point of an operator with spin can then be used as a consistency check that the various structure constants have to satisfy.

As concerns the practical perturbative computation of the two- and three-point functions, I first generate all relevant graphs for the operators without derivatives and then act on them with suitable combinations of them, corresponding to the operators with spin. These can be derived from the explicit definition \eqref{eq:twist2}, expanding the covariant derivatives perturbatively, that is including the contributions from the gauge connections. As a technical detail, the Gegenbauer polynomials have to be considered in $d=4-2\epsilon$ dimensions, for consistency with the rest of the computation which is performed in dimensional regularization, and their coefficients expanded in the regulator up to the required order.
In momentum space the addition of derivatives translates into integrals with higher and higher powers of irreducible numerators consisting of scalar products of the loop momenta with the external null vectors.
As spelled out below in more detail, these can be handled by reduction via integration-by-parts (IBP) identities \cite{Chetyrkin:1981qh,Tkachov:1981wb}.
At one loop this can be performed quite straightforwardly by hand and for generic spins, reducing the only complicated topology to (tensor and with higher powers of the propagators) bubble integrals, which can be computed directly. 
At two loops the reduction becomes more cumbersome and in practice I have resorted to an automated procedure at fixed spins, namely the FIRE5 implementation \cite{Smirnov:2008iw,Smirnov:2013dia,Smirnov:2014hma} (with LiteRed \cite{Lee:2012cn,Lee:2013mka}) of the Laporta algorithm \cite{Laporta:1996mq,Laporta:2001dd}.
The higher the spins the larger the expressions grow and the longer the IBP reduction takes. This is the reason why I capped my two-loop computation to low values of the operators spins.

In the three-point function calculation, I remark that integrating over one insertion point introduces additional powers of propagators if the operator inserted at that point consists of two fields (as is mostly the case in this paper), or an extra vertex otherwise. Both cases are treated as a two-point function problem, by reduction to master integrals via IBP identities.

\section{One loop structure constants}\label{sec:1loop}

I consider three-point functions with two twist-two operators with even spins $j_1$ and $j_2$ and a third protected operator. 
This could be for instance a pair of twist-two operators in the $sl(2)$ sector as in \eqref{eq:twist2} (and its conjugate) and the third operator a rotated version of a length-two BPS operator, so that a non-trivial contraction at tree level exists. Equivalently, one could consider instead the operators, e.g. $\Tr(X  Y)$, $\Tr(\bar X \bar Z)$ and $\Tr(\bar Y Z)$ and apply covariant derivatives on two of them.  
After some computation, the tree level structure constants evaluate
\begin{equation}\label{eq:structtree}
{\cal C}_l^{(0)} = (-1)^{l}\, \frac{ 2^{j_1+j_2-l}\, \Gamma^2 \left(j_1+1\right) \Gamma^2 \left(j_2+1\right)}{\Gamma^2 (l+1)\, \Gamma \left(j_1-l+1\right) \Gamma \left(j_2-l+1\right)}
\end{equation}
These numbers depend on the normalizations of the operators, that in the formula above I have chosen in such a way that the structure constant is unity for the scalar case ($j_1=j_2=l=0$). In any case, the ratios between the various ${\cal C}_l$ components are completely fixed by the general formula \eqref{eq:structtree} and that is its main significance. As in \eqref{eq:3ptstructure}, it is understood that $l\leq\text{min}(j_1,j_2)$ and somehow consistently the structure constants in \eqref{eq:structtree} vanish otherwise, due to the gamma functions in the denominator becoming singular.

I compute the one-loop corrections to the structure constants, using the method outlined in section \ref{sec:method}.
The number of the relevant Feynman diagrams is small, however each features two separate finite sums with ranges set by the spins of the unprotected operators, according to their explicit definition \eqref{eq:twist2}.
At one-loop order such sums can be single or double at most, the latter case occurring when one gauge field from the covariant derivatives is selected in \eqref{eq:twist2}.
The indices of such sums appear as powers of loop momenta contracted with the null vectors $z_1$ and $z_2$ in the numerators. Such integrals are two-loop in momentum space, where the most complicated topology is a kite.
Using an integration-by-parts identity, the latter can be reduced to sums of integrals of lower topologies, which are expressible in terms of bubble integrals. The latter can be finally evaluated explicitly for generic tensor structures, in terms of single or double sums over scalar bubble topologies.
The main message is that the diagrams can be evaluated explicitly, however they depend on several sums and I was not able to derive a general expression for arbitrary spins (though it might be doable). Moreover, the complexity of such an evaluation increases quickly with the spins.

Alternatively, giving up a solution in full generality, a more efficient approach at fixed spins, consists in feeding directly an IBP solver on the market (I have used FIRE5 \cite{Smirnov:2014hma}, for instance) with all the relevant integrals and then evaluate them in terms of the two master integrals of the problem. 
Of course also such a reduction increases in complexity with the spins, but it still allows to evaluate quickly the problem for reasonably high values (order 20).

After the diagrams are solved and a Fourier transform is taken, one has to re-normalize the unprotected operators and in particular take into account mixing.
Instead of computing the two-point function explicitly, I benefited from previously derived results for arbitrary spin \cite{Plefka:2012rd,Belitsky:2007jp}.
In particular, the first entry on the diagonal of the re-normalization matrix \eqref{eq:renormalization} (the relevant part for the current computation) reads for arbitrary spin $j$ (and in a suitable scheme removing some trivial factors to avoid clutter)
\begin{equation}
Z_{j} = 1 + \left(\frac{4\, S_1(j)}{\epsilon } - 6 S_1^2(j) + 4 S_1(j) S_1(2 j)\right)\lambda + O(\epsilon)+O(\lambda^2) 
\end{equation}
and the terms giving rise to mixing with descendants of spin $k$ primaries are governed by the formula
\begin{align}
B_{j,k} &= \frac{4\, \left((-1)^{j-k}+1\right)\, \left(2k+1\right)}{(j-k) (j+k+1)}\,   \left(S_1\left(\tfrac{j-k}{2}\right) - S_1\left(\tfrac{j+k}{2}\right) - 2 S_1(j-k) + 2 S_1(j)\right)  \lambda\nonumber\\&~~~~~~~~~~~~~~~~~~~ + O(\epsilon)+O(\lambda^2) 
\end{align}
and the mixing itself involves only tree level three-point functions, which are given by \eqref{eq:structtree}.

Finally, I retrieved the structure constants using \eqref{eq:integrated} and analysed the data.
The outcome of such a study is the following conjecture for the one-loop structure constants
\begin{align}\label{eq:formula}
\frac{{\cal C}_l^{(1)}}{{\cal C}_l^{(0)}} &= -4(j_1+1) (j_2+1) \sum _{i=1}^l\,  \frac{\Gamma^2(i)}{\Gamma^2 (l+1)}\,a^l_i\, \frac{ \Gamma (j_1-i+1) \Gamma (j_2-i+1) \Gamma (j_1+j_2+2)}{\Gamma (j_1+1) \Gamma (j_2+1) \Gamma (j_1+j_2-i+3)}\nonumber\\&
+4 \sum _{i=1}^l
\frac{\Gamma (l+1)}{i\,\Gamma (l-i+1)}\, \frac{\Gamma (j_1-i+1) \Gamma (j_2-i+1) \Gamma (j_1+j_2+2)}{\Gamma (j_1+1) \Gamma (j_2+1)  \Gamma (j_1+j_2-i+2)}
 \left( S_1(j_1) + S_1(j_2) \right) \nonumber\\&
+ 4 S_1^2(j_1) - 4 S_{1}(2 j_1) S_{1}(j_1) - 2 S_2(j_1) + 4 S_{1}^2(j_2) - 4 S_{1}(2j_2) S_{1}(j_2) - 2 S_2(j_2)
\end{align}
where it is understood that $l\leq \text{min}(j_1,j_2)$, otherwise the correspondent $l$th structure constant vanishes identically.
This conjecture is based mainly on intuition derived from the steps of the perturbative computations and the functions it involves. But let me stress that I have not derived the formula, I have guessed it and I have checked it extensively against all the structure constants that I could compute perturbatively. And it passed the tests.

The dependence on the spins occurs in three qualitatively different manners.
The first consists of harmonic sums of the spins with degree of transcendentality two, which disentangle between the two operators and are basically two copies of the terms appearing in the structure constant of a twist-two operator with spin and two protected operators \cite{Eden:2012rr}. The second term features single harmonic sums of the spins, whose coefficient is a symmetric rational function of the spins that I determined completely. In \eqref{eq:formula} it is expressed as a sum over an index that ranges over the tensor structure label of the structure constant. The sum can be also expressed as an hypergeometric function $_4F_3(1,1,-j_1-j_2,1-l;2,1-j_1,1-j_2;1)$ with some pre-factors (for $1\leq l\leq \text{min}(j_1,j_2)$), for hypergeometric functions lovers.
Finally, a rational function of the spins appears, which in \eqref{eq:formula} is expressed again as a sum. This form constrains significantly the expression of such a rational function and in particular leaves just $l$ undetermined parameters for ${\cal C}_l$. What I mean by constraining is that one could also have cooked up an ansatz for this piece consisting of a rational function with a fixed denominator (which is easy to infer as factorial powers of the spins) and a symmetric polynomial of $j_1$ and $j_2$ whose degree increases with $l$.
Then, the number of unfixed coefficients of such a polynomial would grow as $\frac{l(l+1)}{2}$, instead of $l$.

I have not been able to infer a general form for such coefficients, which would grant a complete (albeit a bit conjectural) knowledge of all one-loop structure constants for the three-point functions under exam. Likewise, I am not in a position to exclude that a more compact or enlightening expression for these rational functions exists, that would expose more clearly the nature of such numbers. 
Still, the undetermined coefficients for ${\cal C}_l$, can be fixed by computing explicitly a set of $l$ independent three-point functions and then \eqref{eq:formula} determines uniquely the given structure constant for all $j_1$ and $j_2$.

Explicitly, I have computed the values for the coefficients $a_{i}^l$ up to $l=14$, which I report in table \ref{tab:as}.
\afterpage{
\begin{landscape}
\begin{table}
\small
\begin{tabular}{|l|llllllllll}
\hline
\diagbox{i}{l} & 1 & 2 & 3 & 4 & 5 & 6 & 7 & 8 & 9 & 10 \\
\hline
1 & 1 & 7 & 85 & 1660 & 48076 & 1942416 & 104587344 & 7245893376 & 628308907776 &
66687811660800 \\
2 & & 5 & 113 & 3140 & 116324 & 5678064 & 356451696 & 28101521664 & 2724406182144 & 
319004276659200 \\
3& & & 20 & 1094 & 60038 & 3867768 & 300911832 & 28258144128 & 3175649048448 & 
422543592230400 \\
4& & & & 94 & 10254 & 985944 & 101850456 & 11913446784 & 1601668439424 &
247962089203200 \\
5& & & & & 524 & 100464 & 15528336 & 2416655904 & 405380065824 & 75189548419200 \\
6& & & & & & 3408 & 1051728 & 245164128 & 54764834400 & 12683323785600 \\
7& & & & & & & 25416 & 11836656 & 3962480112 & 1222612862400 \\
8& & & & & & & & 214128 & 143262576 & 66263716800 \\
9& & & & & & & & & 2012832 & 1861070400 \\
10& & & & & & & & & & 20894400 \\
\hline
\end{tabular}


\begin{tabular}{l|llll}
\hline
\diagbox{i}{l} & 11 & 12 & 13 & 14 \\
\hline
1 & 8506654697548800 & 1284292319599411200 & 226530955276874956800 & 46165213716463676620800 \\
2 & 44422313180083200 & 7259809894343884800 & 1376527546436209459200 & 299803496793988600627200 \\
3&  65858628437798400 & 11904037388242329600 & 2472451846202358374400 & 585198645069048591974400 \\
4&  44063172266227200 & 8940760300982476800 & 2059299390621783859200 & 535201352007492125491200 \\
5&  15565939686403200 & 3605052787707340800 & 933058622012068915200 & 269159343300345841459200 \\
6&  3147840251913600 & 849807651199334400 & 251172231200228505600 & 81454941449147995545600 \\
7&  379027212350400 & 122711161415961600 & 42289585295028710400 & 15665547213313430630400 \\
8&  27375173092800 & 11073038073523200 & 4577250012251212800 & 1977372395519683276800 \\
9&  1152769118400 & 621467688729600 & 321004318102502400 & 166354530738194534400 \\
10& 25874942400 & 20917207065600 & 14401428443366400 & 9326569479652454400 \\
11& 237458880 & 383814305280 & 396282211591680 & 342105394966272000 \\
12& & 2932968960 & 6055165877760 & 7839464449259520 \\
13& & & 39126516480 & 101294501736960 \\
14& & & & 560704273920 \\
\end{tabular}
\caption{Table of the coefficients $a_i^l$ of \eqref{eq:formula}, with $i$ running vertically and $l$ horizontally.\label{tab:as}}
\end{table}
\end{landscape}}
The normalization $\frac{1}{\Gamma^2(l+1)}$ that I chose in \eqref{eq:formula}, appears to remove all denominators and I conjecture that the resulting coefficients  $a_{i}^l$ are integer numbers. This also leads me to infer that they are given by combinations of at most double sums with upper bound $l$.
In particular I could only find a pattern for the edges of the triangle in table \ref{tab:as} which I conjecture to be given by
\begin{align}
a^l_1 &= \frac{\Gamma^2(l+1)}{2} \left( S_1^2(l) + S_2(l) \right)\\
a^l_{l} &= \Gamma(l+1)\left( S_1(l-1) + S_1(l)\right)
\end{align}
The results in table \ref{tab:as} allow to compute the one-loop structure constants associated to the first 15 tensor structures according to \eqref{eq:3ptstructure} for all spins. In particular, they fix the one-loop three-point functions completely for one unprotected operator with up to 14 units of spin and another unprotected operator with generic spin $j$.
Results with higher spin can in principle be derived by computing explicitly more three-point functions as described above, however the procedure is doomed to become more and more time consuming. Hence, a completely analytic in spins solution would be highly preferable, but, as I said, I have not been able to perform such an evaluation yet.

\paragraph{Structure constants with a scalar Konishi}
For completeness I consider here the case of three-point functions of twist-two operators with two unprotected ones, where one is a spinning $sl(2)$ sector operator of the form \eqref{eq:konishi} and the other is a scalar Konishi R-symmetry singlet \eqref{eq:konishi}.

In such a case the tensor structure allows for a single structure constant and moreover having a single operator with spin allows for significant simplifications, as studied in \cite{Plefka:2012rd}.
Therefore the one-loop correction to such a structure constant can be computed analytically for any spin $j$.
Precisely, it can be given compactly as a difference with the corresponding structure constant with one spinning and two BPS operators, which I quote from \cite{Plefka:2012rd,Eden:2012rr,Dolan:2004iy}
\begin{equation}
\frac{{\cal C}_{0,j,BPS}}{{\cal C}^{(0)}_{0,j,BPS}} = 1 + \left( 4 S_1^2(j) - 4 S_1(j)S_1(2 j) - 
  2 S_2(j)\right)\lambda + O(\lambda^2)
\end{equation}
and it reads
\begin{equation}
\frac{{\cal C}_{0,j,K}}{{\cal C}^{(0)}_{0,j,K}} - \frac{{\cal C}_{0,j,BPS}}{{\cal C}^{(0)}_{0,j,BPS}} = 6 \lambda  \left(S_1(j)-1\right) + O(\lambda^2)
\end{equation}

\section{Two-loop structure constants}

The quantum corrections to the structure constants can be computed, using the method described previously, also at higher loop level.
In particular, I have evaluated explicitly the relevant diagrams for the two-loop correction to the structure constants.
In momentum space they look like three-loop two-point functions with powers of the loop momenta contracted with the null vectors in the numerators.
Needless to say, the computation is decidedly more cumbersome than at one loop. Apart from the higher number of diagrams and the more complicated expansion of the operators \eqref{eq:twist2} at two loops, some additional technical complications arise.
In particular, the re-normalization of the operators and the mixing matrix (including finite corrections and subleading terms in the dimensional regularization parameter at lower perturbative order) have to be computed explicitly, as I did not find general results for them in the literature.
Moreover, the reduction of three-loop tensor integrals is much more complicated and I resorted to an automated process with FIRE5, and limited the analysis to low values of the spins.

After many IBP reductions and taking into account mixing, with the notation of \eqref{eq:3ptstructure}, I have obtained (for the ratios of the structure constants by their tree level value \eqref{eq:structtree})
\begin{equation}
{\cal C}_{2,2,BPS} \rightarrow \left\{
\begin{array}{l}\displaystyle
{\cal C}_0/{\cal C}_0^{(0)} = 1-12 \lambda+147 \lambda ^2\\[4mm]\displaystyle
{\cal C}_1/{\cal C}_1^{(0)} = 1-6 \lambda + \frac{111}{2}\lambda ^2\\[4mm]\displaystyle
{\cal C}_2/{\cal C}_2^{(0)} = 1+6 \lambda -87 \lambda ^2
\end{array}\right.
\end{equation}

\begin{equation}
{\cal C}_{2,4,BPS} \rightarrow \left\{
\begin{array}{l}\displaystyle
{\cal C}_0/{\cal C}_0^{(0)} = 1-\frac{1781}{126} \lambda+\left(\frac{2712265}{15876}+14 \zeta (3)\right) \lambda ^2\\[4mm]\displaystyle
{\cal C}_1/{\cal C}_1^{(0)} = 1-\frac{4583}{504} \lambda+\left(\frac{2718197}{31752}+14 \zeta (3)\right) \lambda ^2\\[4mm]\displaystyle
{\cal C}_2/{\cal C}_2^{(0)} = 1+\frac{65}{63} \lambda+\left(-\frac{474107}{15876}+14 \zeta (3)\right) \lambda ^2
\end{array}\right.
\end{equation}

\begin{equation}
{\cal C}_{2,6,BPS} \rightarrow \left\{
\begin{array}{l}\displaystyle
{\cal C}_0/{\cal C}_0^{(0)} = 1-\frac{12692 }{825} \lambda+\left(\frac{246135733}{1306800}+\frac{114 \zeta (3)}{5}\right) \lambda ^2\\[4mm]\displaystyle
{\cal C}_1/{\cal C}_1^{(0)} = 1-\frac{34763 }{3300} \lambda+\left(\frac{673255007}{6534000}+\frac{114 \zeta (3)}{5}\right) \lambda ^2\\[4mm]\displaystyle
{\cal C}_2/{\cal C}_2^{(0)} = 1-\frac{239 }{330} \lambda+\left(-\frac{219497639}{32670000}+\frac{114 \zeta (3)}{5}\right) \lambda ^2
\end{array}\right.
\end{equation}

\begin{equation}
{\cal C}_{4,4,BPS} \rightarrow \left\{
\begin{array}{l}\displaystyle
{\cal C}_0/{\cal C}_0^{(0)} = 1-\frac{1025}{63} \lambda+\frac{14378795}{63504} \lambda ^2\\[4mm]\displaystyle
{\cal C}_1/{\cal C}_1^{(0)} = 1-\frac{6625}{504} \lambda+\frac{20988115}{127008} \lambda ^2\\[4mm]\displaystyle
{\cal C}_2/{\cal C}_2^{(0)} = 1-\frac{500}{63} \lambda+\frac{9858115}{127008} \lambda ^2\\[4mm]\displaystyle
{\cal C}_3/{\cal C}_3^{(0)} = 1+\frac{325}{84} \lambda-\frac{9823085}{127008} \lambda ^2\\[4mm]\displaystyle
{\cal C}_4/{\cal C}_4^{(0)} = 1+\frac{725}{14} \lambda-\frac{33859255}{63504} \lambda ^2
\end{array}\right.
\end{equation}

\begin{equation}
{\cal C}_{4,6,BPS} \rightarrow \left\{
\begin{array}{l}\displaystyle
{\cal C}_0/{\cal C}_0^{(0)} = 1-\frac{607039}{34650} \lambda+\left(\frac{2916214006}{12006225}+\frac{44 \zeta (3)}{5}\right) \lambda ^2\\[4mm]\displaystyle
{\cal C}_1/{\cal C}_1^{(0)} = 1-\frac{86864}{5775} \lambda+\left(\frac{368404919191}{1920996000}+\frac{44 \zeta (3)}{5}\right) \lambda ^2\\[4mm]\displaystyle
{\cal C}_2/{\cal C}_2^{(0)} = 1-\frac{1134277}{103950} \lambda+\left(\frac{55429699999}{480249000}+\frac{44 \zeta (3)}{5}\right) \lambda ^2\\[4mm]\displaystyle
{\cal C}_3/{\cal C}_3^{(0)} = 1-\frac{178813}{69300} \lambda+\left(-\frac{38694354697}{3841992000}+\frac{44 \zeta (3)}{5}\right) \lambda ^2\\[4mm]\displaystyle
{\cal C}_4/{\cal C}_4^{(0)} = 1+\frac{816691}{34650} \lambda+\left(-\frac{15231046612}{60031125}+\frac{44 \zeta (3)}{5}\right) \lambda ^2
\end{array}\right.
\end{equation}
Heuristically, the coefficient of the $\zeta(3)$ term at two loops can be conjectured to be in general $24|S_1(j_1)-S_1(j_2)|$ for all structure constants of a given three-point function.
In other words, this term is proportional to the difference between the one-loop anomalous dimensions of the operators.
From the one-loop analysis it is reasonable to presume that the other pieces could be given by a suitable combination of the harmonic sums appearing in the situation with only one unprotected operator \cite{Eden:2012rr}. 
However, the cases hitherto analyzed did not provide me enough data to identify a pattern to formulate an ansatz. 
Moreover, I expect to need many more data to be able to do so, since the number of different harmonic sums that can appear in the result is conceivably much larger than at one loop. Regrettably, as I said, a computational bottleneck in the reduction of the integrals required for the two-loop corrections kicks in pretty soon when raising the spins.
Perhaps one could address this issue by trying to solve the relevant IBP reductions suited to the particular problem at hand, along the lines of \cite{Kosower:2018obg}, instead of using the available software on the market, but I have not attempted such an approach yet.

Anyway, the results above constitute some solid perturbative data that can be used, for instance, for consistency checks with an integrability based computation of three-point functions or in the OPE analysis of higher-point correlators (for which integrability can also provide powerful tools \cite{Fleury:2016ykk,Basso:2017khq,Eden:2016xvg}).

\paragraph{Structure constants with a scalar Konishi}
Again, I conclude with some results for structure constants when only one unprotected operator has spin and the other is a scalar Konishi (or a BPS operator itself, instead).
I report the explicit results for spin up to six in table \ref{tab:Konishi}.
\begin{table}
\begin{tabular}{|c|c|c|}
\hline
\rule{0pt}{4ex}$j$ & $\displaystyle 1- \frac{{\cal C}_{0,j,BPS}}{{\cal C}^{(0)}_{0,j,BPS}}$ & $\displaystyle 1-\frac{{\cal C}_{0,j,K}}{{\cal C}^{(0)}_{0,j,K}}$ \rule[-3.5ex]{0pt}{0pt} \\[2mm]\hline
\rule{0pt}{4ex}$0$ & $0$ & $6 \lambda -\left(66 +36 \zeta (3)\right)\lambda ^2$ \\[2mm]
\rule{0pt}{1ex}$2$ & $6 \lambda -\left(66 +36 \zeta (3)\right)\lambda ^2$ & $3 \lambda -21 \lambda ^2$ \\[2mm]
\rule{0pt}{4ex}$4$ & $\displaystyle \frac{1025 \lambda }{126}-\left(\frac{3532955}{31752}+50 \zeta (3)\right) \lambda ^2$ & $\displaystyle \frac{103 \lambda }{63}-\left(\frac{28465}{15876}+14 \zeta (3)\right) \lambda ^2$ \rule[-3ex]{0pt}{0pt} \\[2mm]
\rule{0pt}{4ex}$6$ & $\displaystyle \frac{7742 \lambda }{825}-\left(\frac{189088963}{1306800}+\frac{294 \zeta (3)}{5}\right) \lambda ^2$ & $\displaystyle \frac{1129 \lambda }{1650}-\left(-\frac{26487091}{6534000}+\frac{114 \zeta (3)}{5}\right) \lambda ^2$\rule[-3ex]{0pt}{0pt} \\[2mm]
\hline
\end{tabular}
\caption{Two-loop three-point functions with a spin $j$ operator and a Konishi.\label{tab:Konishi}}
\end{table}
In the first column I am testing my computation against already known results \cite{Eden:2012rr}, which in particular have been derived for generic spin. The second column provides novel results, where one of the BPS operators is replaced by an unprotected scalar Konishi singlet.
It is a safe bet that the term proportional to $\zeta(3)$ is determined to all spins to be
$12(2S_1(j)-3)$, with $j\geq 2$, which is again proportional to the difference of the anomalous dimensions at one loop (alike the case with two operators with spin). It would be interesting to ascertain whether such a pattern persists in some fashion at higher loops.
I have not been able to find the close form of the rational term at two loops, which would presumably be given again by a combination of various harmonic sums of mixed transcendentality (as is the case at one loop).

\acknowledgments

This work has been supported by DFF-FNU through grant number DFF-4002-00037.
I would like to thank Jan Ambjorn for letting me use his computing facilities at NBI, which sped up considerably the computation of the results in this paper.

\vfill
\newpage

\appendix

\section{More one-loop structure constants}

In this appendix I collect some of the explicit results for the structure constants at one loop, at fixed values of the spins of the operators, which have been used to derive and test the conjecture \eqref{eq:formula} and are not covered by the coefficients of table \ref{tab:as}.
\begin{align}
\left.\frac{{\cal C}_{15}^{(1)}}{{\cal C}_{15}^{(0)}}\, \right|_{j_1=16, j_2 =16} &= \frac{6481018936656560253511}{21405703134816000} \nonumber\\
\left.\frac{{\cal C}_{15}^{(1)}}{{\cal C}_{15}^{(0)}}\, \right|_{j_1=18, j_2 =16} &= \frac{2654063224812622326609799}{68048730265580064000}\nonumber\\
\left.\frac{{\cal C}_{15}^{(1)}}{{\cal C}_{15}^{(0)}}\, \right|_{j_1=18, j_2 =18} &= \frac{52518437784893710046987}{10147617671182992000}\nonumber\\
\left.\frac{{\cal C}_{15}^{(1)}}{{\cal C}_{15}^{(0)}}\, \right|_{j_1=20, j_2 =16} &= \frac{6274359527119399790751221803}{621897345897136204896000} \nonumber\\
\left.\frac{{\cal C}_{15}^{(1)}}{{\cal C}_{15}^{(0)}}\, \right|_{j_1=20, j_2 =18} &= \frac{5179111631301167954418495997}{3524084960083771827744000} \nonumber\\
\left.\frac{{\cal C}_{15}^{(1)}}{{\cal C}_{15}^{(0)}}\, \right|_{j_1=20, j_2 =20} &= \frac{7285092570367542539344391}{15754732762727450524032} \nonumber\\
\left.\frac{{\cal C}_{16}^{(1)}}{{\cal C}_{16}^{(0)}}\, \right|_{j_1=16, j_2 =16} &= \frac{56130165521942126696117}{2675712891852000}\nonumber\\
\left.\frac{{\cal C}_{16}^{(1)}}{{\cal C}_{16}^{(0)}}\, \right|_{j_1=18, j_2 =16} &= \frac{13038732103616718985565903}{8506091283197508000}\nonumber\\
\left.\frac{{\cal C}_{16}^{(1)}}{{\cal C}_{16}^{(0)}}\, \right|_{j_1=18, j_2 =18} &= \frac{602866752681361460628109}{7610713253387244000}\nonumber\\
\left.\frac{{\cal C}_{16}^{(1)}}{{\cal C}_{16}^{(0)}}\, \right|_{j_1=20, j_2 =16} &= 
\frac{21187697505559626222458936441}{77737168237142025612000}\nonumber\\
\left.\frac{{\cal C}_{16}^{(1)}}{{\cal C}_{16}^{(0)}}\, \right|_{j_1=20, j_2 =18} &= \frac{6246558271675934507704152059}{440510620010471478468000} \nonumber\\
\left.\frac{{\cal C}_{16}^{(1)}}{{\cal C}_{16}^{(0)}}\, \right|_{j_1=20, j_2 =20} &= \frac{4623044287691782810404245329}{1673940356039791618178400} \nonumber\\
\left.\frac{{\cal C}_{17}^{(1)}}{{\cal C}_{17}^{(0)}}\, \right|_{j_1=18, j_2 =18} &= \frac{5534098436350486058537}{2025742148892000}\nonumber\\
\left.\frac{{\cal C}_{17}^{(1)}}{{\cal C}_{17}^{(0)}}\, \right|_{j_1=20, j_2 =18} &= \frac{429533698548664859661118331}{1524258200728275012000} \nonumber\\
\left.\frac{{\cal C}_{17}^{(1)}}{{\cal C}_{17}^{(0)}}\, \right|_{j_1=20, j_2 =20} &= \frac{168525193326165153092498713}{5792181162767445045600} \nonumber\\
\left.\frac{{\cal C}_{18}^{(1)}}{{\cal C}_{18}^{(0)}}\, \right|_{j_1=18, j_2 =18} &= \frac{42459559466946840172477}{172121881932000}\nonumber\\
\left.\frac{{\cal C}_{18}^{(1)}}{{\cal C}_{18}^{(0)}}\, \right|_{j_1=20, j_2 =18} &= \frac{7412499710616596349975089777}{508086066909425004000} \nonumber\\
\left.\frac{{\cal C}_{18}^{(1)}}{{\cal C}_{18}^{(0)}}\, \right|_{j_1=20, j_2 =20} &= \frac{67218719880291136167753563}{113572179662106765600} \nonumber\\
\left.\frac{{\cal C}_{19}^{(1)}}{{\cal C}_{19}^{(0)}}\, \right|_{j_1=20, j_2 =20} &= \frac{331549674578564713855823}{12584175031812384} \nonumber\\
\left.\frac{{\cal C}_{20}^{(1)}}{{\cal C}_{20}^{(0)}}\, \right|_{j_1=20, j_2 =20} &= \frac{33799035747820846186865869}{11259525028463712} \nonumber
\end{align}

\bibliographystyle{JHEP}

\bibliography{biblio2}

\end{document}